\documentclass[10pt]{bmc_article}    

\pdfoutput=1

\usepackage{cite} 
\usepackage{url}  
\usepackage{ifthen}  
\usepackage{multicol}   
\usepackage[utf8]{inputenc} 

\usepackage[english]{babel}
\usepackage{amsmath}
\usepackage{amssymb}
\usepackage{amsfonts}
\usepackage{booktabs}
\usepackage{multirow}
\usepackage{tabularx}
\usepackage{graphicx}

\newcounter{specialsubsectionfigure}

\newcommand{\specialsubsectionfigure}[1]{%
\refstepcounter{specialsubsectionfigure}%
\subsection*{#1}}

\newcounter{specialsubsectiontable}

\newcommand{\specialsubsectiontable}[1]{%
\refstepcounter{specialsubsectiontable}%
\subsection*{#1}}

\newboolean{publ}

\newenvironment{bmcformat}{\fussy\setboolean{publ}{true}}{\fussy}

\begin{document}
\begin{bmcformat}


\title{Laser-induced etching of few-layer graphene synthesized by Rapid-Chemical Vapour Deposition on Cu thin films}
 

\author{Marco Piazzi\correspondingauthor$^{1,2}$%
         \email{Marco Piazzi\correspondingauthor - m.piazzi@inrim.it}
       \and 
         Luca Croin$^{1,3}$%
         \email{Luca Croin - l.croin@inrim.it}%
       \and 
         Ettore Vittone$^2$%
         \email{Ettore Vittone - ettore.vittone@unito.it}
       and 
         Giampiero Amato$^1$%
         \email{Giampiero Amato - g.amato@inrim.it}%
      }


\address{%
    \iid(1)Quantum Research Laboratory, Istituto Nazionale di Ricerca Metrologica, Strada delle Cacce 91, 10135 Turin, Italy\\
    \iid(2)Department of Physics, NIS Centre of Excellence and CNISM, University of Turin, Via Pietro Giuria 1, 10125 Turin, Italy\\
    \iid(3)Department of Applied Science and Technology, Politecnico of Turin, Corso Duca deli Abruzzi 24, 10129 Turin, Italy
}%

\maketitle


\begin{abstract}
The outstanding electrical and mechanical properties of graphene make it very attractive for several applications, Nanoelectronics above all. However a reproducible and non destructive way to produce high quality, large-scale area, single layer graphene sheets is still lacking. Chemical Vapour Deposition of graphene on Cu catalytic thin films represents a promising method to reach this goal, because of the low temperatures ($T < 900$ $^\circ$C) involved during the process and of the theoretically expected monolayer self-limiting growth. On the contrary such self-limiting growth is not commonly observed in experiments, thus making the development of techniques allowing for a better control of graphene growth highly desirable. Here we report about the local ablation effect, arising in Raman analysis, due to the heat transfer induced by the laser incident beam onto the graphene sample.


\end{abstract}

\ifthenelse{\boolean{publ}}{\begin{multicols}{2}}{}


\section*{Background}
Graphene (a single bidimensional layer of carbon atoms arranged in an hexagonal lattice) has attracted a major interest in the last few years because of its astonishing electrical\cite{castroneto:graph_rev, peres:graph_rev,peres:graph_rev_2}, mechanical\cite{lee:mech_prop_graph} and chemical properties\cite{elias:chemic_prop_graph,wang:chemic_prop_graph}, that makes it a good candidate for the future development of nanoelectronics devices. Although the main properties of this material are nowadays well known from a theoretical point of view, an efficient and highly reproducible method to grow high quality, large-scale area, single layer graphene films, suitable for practical applications, is still lacking. For this reason, several techniques have been developed in the last years in order to achieve this goal: the most important are the epitaxial growth of graphene by thermal sublimation of SiC\cite{hass:graph_SiC,deheer:graph_SiC,varchon:graph_SiC,emtsev:graph_SiC,sprinkle:graph_SiC}, the Chemical Vapour Deposition (CVD) synthesis of graphene on various metal catalysts\cite{reina:CVD,nandamuri:CVD,liu:CVD_Ni,somani:CVD_Ni,kim:CVD_Ni,li:CVD_NiCucomparison,li:CVD_Cu,tao:CVD_Cufilmandfoil,lee:CVD_NiCu,kim:CVD_activpathCu} and the chemical reduction of graphene oxide\cite{reduct_graphox_1,reduct_graphox_2,reduct_graphox_3,reduct_graphox_4}. Among these, CVD technique seems to be one of the most promising methods because of the reported possibility\cite{liu:CVD_Ni} of obtaining highly uniform, defect-free graphene flakes as large as $\sim100$ $\mu$m$^2$ in a reproducible, highly accessible and inexpensive way.   

Since CVD synthesis needs a catalyst to activate the chemical decomposition of the carbon precursor (usually methane or ethylene) used for graphene growth at low temperatures ($T<1000$ $^\circ$C), the use of many metals (Ir\cite{coraux:CVD_Ir}, Ru\cite{martoccia:CVD_Ru}, Pt\cite{sasaki:CVD_Pt,starr:CVD_Pt}, Fe\cite{kondo:CVD_Fe}, Ag\cite{di:CVD_Ag}, Ni\cite{liu:CVD_Ni,kim:CVD_Ni,obraztsov:CVD_Ni}, Cu\cite{li:CVD_Cu,tao:CVD_Cufilmandfoil, bae:CVD_Cu}) as catalysts during the process has been reported in literature. Cu is one of the most promising catalyst\cite{mattevi:CVD_Cu} because of the low C solid solubility in it ($0.001-0.008$ weight \% $\sim 1084$ $^\circ$C): this property brings to the formation of only soft (not covalent) bonds between the $\pi$ electrons of $2$p$_z$ orbitals of sp$^2$ hybridized C atoms and the $4$s electrons of Cu, without formation of any carbide phase during the growth process. As a consequence, formation of graphene should stop after one single layer has been formed: this makes CVD growth of graphene on Cu very attractive. Nonetheless, many experiments show that actually such a self-limiting behaviour is hardly observed, since few-layered graphitic structures are usually grown on Cu substrates.    

For this reason, to obtain graphene monolayer, several post processing techniques have been proposed to selectively etch atomic graphene layers. Among the various approaches (e.g. heat-induced etching by oxygen\cite{oxigen_etch}, e-beam lithography assisted technique\cite{geimnov:adesive_tape_graph2004,zhang:exper_qhe}, graphene cutting by carbon-soluble metals\cite{cutting_etch,cutting_etch_2}) the thinning of atomic carbon multilayers by laser irradiation\cite{han:laser_etch} can be a promising method to obtain graphene monolayer. In this last work, authors show how the central r\^{o}le in graphene etching is held both by the laser irradiation used for the Confocal Raman Spectroscopy performed on the samples, and by the SiO$_2$/Si substrate on top of which few-layered graphene has been transferred: the heat produced by the irradiation ``burns'' (in presence of oxygen) locally the outermost C layers, while the innermost layer (the one bounded to the substrate) is left unetched because of the presence of the SiO$_2$ layer acting as heat sink.\\
Cu, being a metal, has thermal conductivity higher than SiO$_2$ and can represent therefore an enhanced heat sink, so it can be expected to observe a similar behaviour also for graphene grown by CVD on it.

Here we report the change in shape and position of the G and 2D peaks observed in Raman spectra acquired at different time intervals on the same spot of a graphene sample synthesized by CVD on Cu thin films: the evolution of the spectra, indicating that the structure of graphene is changing during the exposure to the laser used during Raman analysis, is compatible with a decrease in the number of graphene layers present on the substrate. This decrease may be attributed to the same laser-induced etching effect observed on graphene deposited on SiO$_2$/Si substrates. If this is the case, an efficient and easy graphene etching technique can be developed and a promising way to obtain high uniform, large-scale area, monolayer graphene can be envisaged.
 
\section*{Methods}
\subsection*{Cu deposition}
The samples subjected to CVD process have been prepared by e-beam evaporation of $500$ nm copper thin film on top of a p-type (100) oriented Si wafer ($\sim 1$ cm$^2$) with $\sim 300$ nm thermal SiO$_2$. The deposition has been carried out in a load-lock chamber at a base pressure of $\sim 10^{-8}$ mbar and deposition pressure of $\sim 10^{-6}$ mbar, with an average growth rate of $3-5$ \AA/s.\\
The thickness $d$ of deposited Cu film has been chosen in order to limit the known problem\cite{mattevi:CVD_Cu} of dewetting occurring on very thin films ($d < 500$ nm) at temperatures $\gtrsim 800$ $^\circ$C. 

Scanning Electron Microscopy (SEM) images of the samples (taken with a FEI \textit{InspectF} Scanning Electron Microprobe) after Cu deposition show a uniform coverage of the SiO$_2$ surface characterized by a Cu polycrystalline structure with grains of $\sim 90$ nm as typical size (\ref{Cu500noCVD}(a)).  Moreover, Scanning Tunneling Microscopy (STM) scanning an area of  $\sim 10^{-2}$ $\mu$m$^2$ allowed us to evaluate the roughness of the Cu surface (\ref{Cu500noCVD}(b)). The resulting root mean square (RMS) of $\sim 2$ nm (an order of magnitude higher than single layer graphene thickness, $\sim 3.3$ \AA), together with the topographic behaviours obtained for certain scanning directions (\ref{Cu500noCVD}(c)), showing among others height variations as small as few angstroms, makes ineffective the use of Atomic Force Microscopy (AFM) to detect any change in the number of graphene layers eventually present on Cu: the changes in height produced by the latter effect would be hardly distinguishable from topographic changes due to the roughness of the substrate's surface.

\subsection*{CVD growth process}
Before undergoing CVD process, samples have been carefully cleaned in acetone and isopropanol. CVD has been then performed in a Rapid Thermal Annealing (RTA) system (Jipelec \textit{JetFIRST 100}) suitable for depositions in Low Vacuum conditions ($p_\mathrm{min}\sim 10^{-2}$ mbar) up to $T_\mathrm{max}\sim 1300$ $^\circ$C. The system is equipped with four gas lines controlled through mass flow meters and is characterized by a small heat capacity allowing for fast cooling-down processes, up to $\sim 300$ $^\circ$C/min (see \ref{RTAsystem_function}). 

The temperature of the system is controlled by a pyrometer, exposed to the back of the sample holder (a 4'' Si wafer) and calibrated by means of a thermocouple in contact with it (\ref{RTAsystem_function}). The pyrometer sets the power of the lamps. Since the Cu sample undergoing the CVD process is directly exposed to the lamps, the temperature reading by the pyrometer (correct in absence of Cu) is probably lower than the temperature reached by the surface during the process. Therefore, the real deposition temperature may be underestimated. We are performing some studies on this topic in order to solve this ambiguity in the future. However, at the temperature reached during the process as read by the pyrometer dissociation of CH$_4$ and subsequent graphene formation take place, while the undesired Cu dewetting effect is prevented.

\begin{center}
\includegraphics[keepaspectratio,scale=0.37]{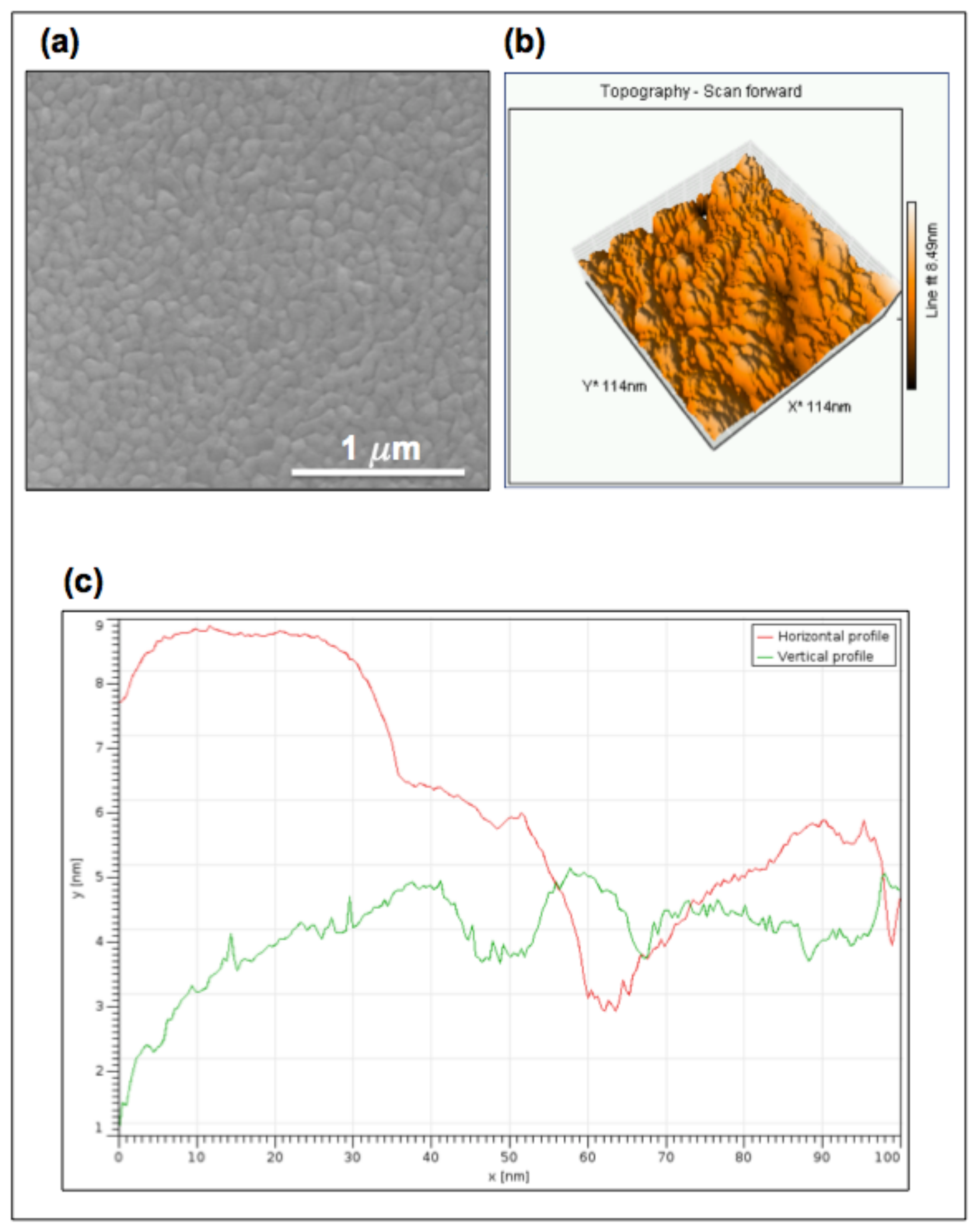}
\end{center}
\specialsubsectionfigure{\footnotesize{Figure 1 - SEM and STM Analyses of e-beam evaporated Cu thin films as deposited}}\label{Cu500noCVD}  
\scriptsize{(a) SEM image of the polycrystalline structure of the Cu surface after e-beam evaporation; (b) STM topographic image of a small portion -$\sim 10^{-2}$ $\mu$m$^{2}$- of the Cu surface; (c) topographic profiles along two directions (red: horizontal, green: vertical) obtained with STM analysis.}
\begin{center}
\includegraphics[keepaspectratio,scale=0.5]{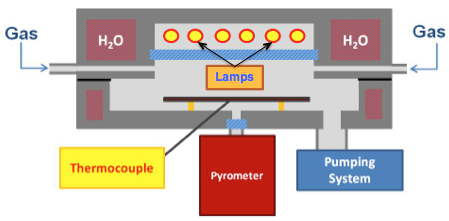}
\end{center}
\specialsubsectionfigure{\footnotesize{Figure 2 - RTA system for graphene deposition}}\label{RTAsystem_function}
\scriptsize{Schematic view of RTA system working principles: heating lamps provide heat only to the sample.}

\normalsize
\vskip 0.5cm
The deposition process started with a fast heating step, increasing the temperature up to 500 $^\circ$C, carried out in vacuum ($p_\mathrm{chamber}=1.2\cdot 10^{-2}$ mbar). A subsequent annealing step (10 minutes long), bringing the system to 725 $^\circ$C, has been then performed under 40 sccm of H$_2$ (p$_\mathrm{chamber}=1.65$ mbar). Purpose of this step is to improve the quality of the Cu film, by increasing the size of the grains present on its surface and, consequently, a low defective growth of graphene due to the reduced number of grain boundaries. The synthesis of graphene has been performed by flowing ultrahigh purity CH$_4$, with a flow rate of 10 sccm, for 5 minutes (p$_\mathrm{chamber}=3.3\cdot 10^{-1}$ mbar), without any H$_2$ flow, following \cite{tao:CVD_Cufilmandfoil} in which H$_2$ has been reported to be detrimental for the final quality of graphene sheets when using Cu thin films. After deposition the system has been cooled-down to room temperature in three steps: a first one (at a rate of $\sim 42$ $^\circ$C/min), down to 450 $^\circ$C in gas-free conditions, a faster one by switching the heaters off (at a rate of $\sim 180$ $^\circ$C/min), till to $\sim 270$ $^\circ$C, again in vacuum conditions  and a final one carried out in nitrogen atmosphere, to purge the system, down to room temperature. The thermal cycle has been reported in \ref{thermalcycle} for clarity.

\begin{center}
\includegraphics[keepaspectratio,scale=0.47]{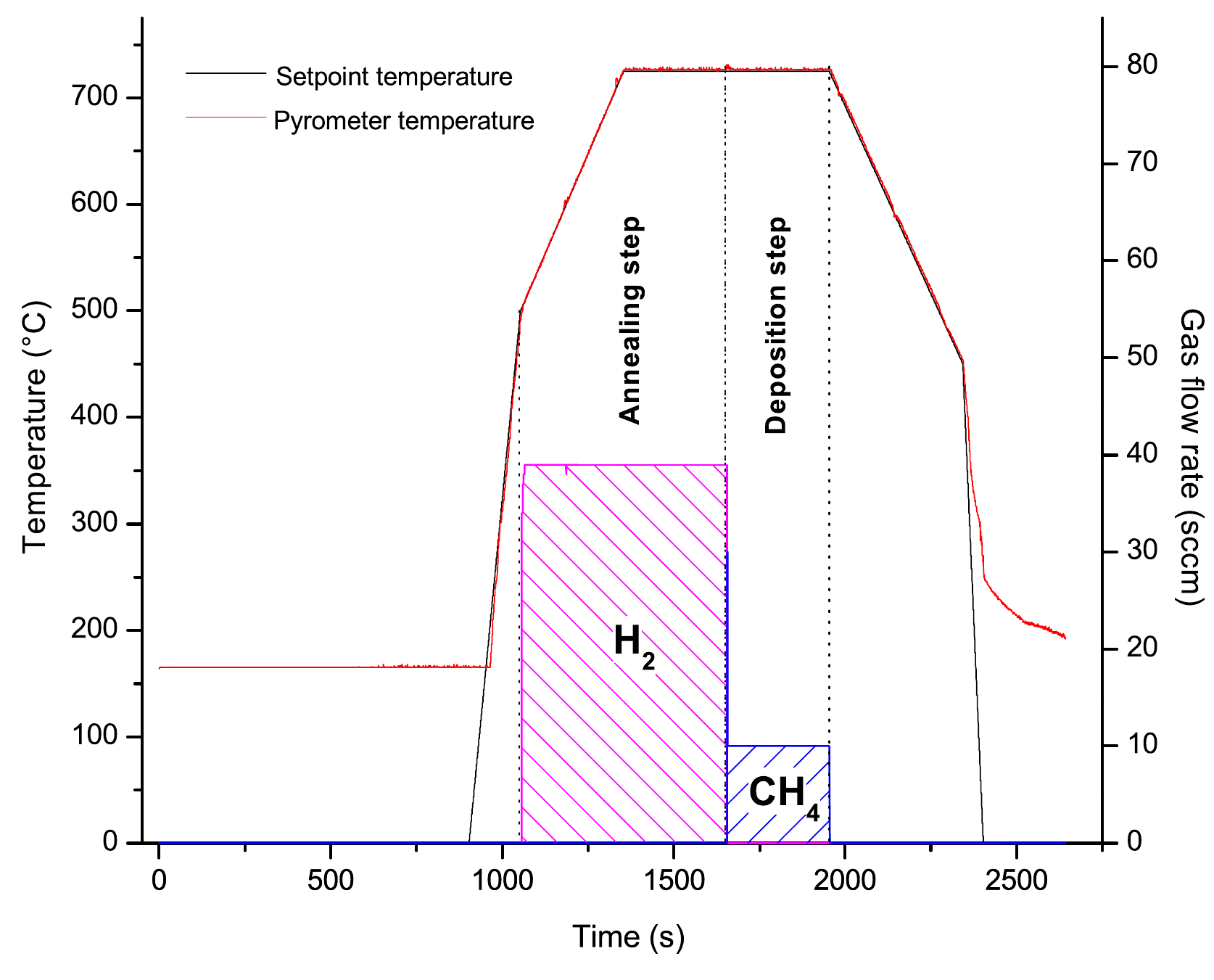}
\end{center}
\specialsubsectionfigure{\footnotesize{Figure 3 - Expected thermal cycle for graphene deposition}}\label{thermalcycle}
\scriptsize{Illustration of the thermal process performed for CVD growth of graphene on top of Cu: see text for details.}
      
\normalsize
\vskip 0.5cm      
\subsection*{Graphene transfer}
Graphene has been transferred from Cu to new insulating (285 nm-SiO$_2$/Si) substrates by means of a standard chemical etching technique. Approximately 1.5 $\mu$m of Poly(methyl-methacrylate) (PMMA) have been firstly deposited on graphene by spinning (at 5000 rpm for 1 minute) and heating (at 165$^\circ$C for 5 minutes) the samples several times. After PMMA deposition, samples have been dipped in a 0.5 M diluted FeCl$_3$ etching solution for $\sim 4$ hours (keeping them at 45$^\circ$C to enhance etching kinetics). Once Cu has been completely etched away from the samples, PMMA/graphene membranes have been picked up onto 285 nm-SiO$_2$/Si substrates and PMMA has been finally removed with acetone.        

\section*{Results and discussion}
Graphene samples have been first characterized by SEM analysis and X-Ray Diffraction (XRD, performed with a Philips \textit{X'Pert Pro} X-ray system using Co radiation beam at 0.8$^\circ$ incident angle) in order to investigate a change in granulometry and crystallographic orientation of the Cu surface, due to high temperature annealing under hydrogen flow.\\ 
SEM image in \ref{Cu500postCVD}(a) shows that Cu surface reorganized itself at the temperature reached during graphene deposition: an increase of grain average size (from $\sim 70$ nm to $\sim 3$ $\mu$m) is observed, as expected. Cu still covers uniformly the sample surface, meaning that the temperature during the annealing and deposition processes was low enough and the growth time short enough to prevent dewetting effects on the catalytic film. Nevertheless, early stage formation of Cu droplets is observed (\ref{Cu500postCVD}(b)).

\begin{center}
\includegraphics[keepaspectratio,scale=0.4]{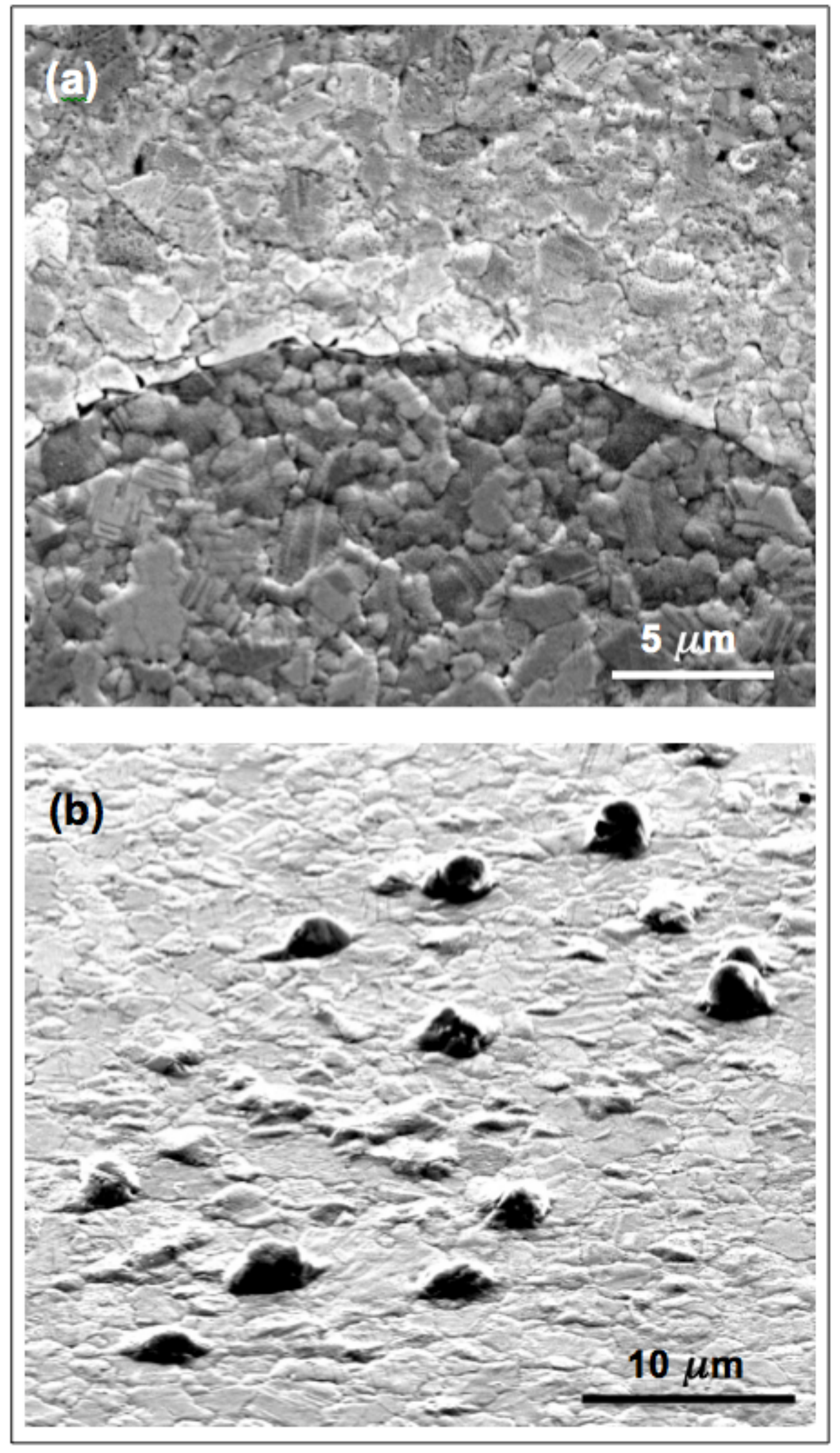}
\end{center}
\specialsubsectionfigure{\footnotesize{Figure 4 - SEM Analysis of Cu samples after graphene deposition}}\label{Cu500postCVD}
\scriptsize{SEM images of a Cu sample after CVD process: (a) Cu surface reorganizes itself by increasing the size of the grains; (b) Cu droplets formation starts to appear on top of the surface.}

\normalsize
\vskip 0.5cm
XRD analysis performed on our samples, in contrast to what observed in other works\cite{tao:CVD_Cufilmandfoil}, shows (\ref{XRD}) changes in the preferential crystallographic orientation of the grains present on the sample surface after CVD process at the nominal temperature of 725 $^\circ$C, involving both (111) and (220) directions, with a decrease in the I$_{(111)}$/I$_{(220)}$ intensity ratio after the thermal treatment. This fact can probably be ascribed to the fast cooling-down rate chosen for the experiment.

\begin{center}
\includegraphics[keepaspectratio,scale=0.4]{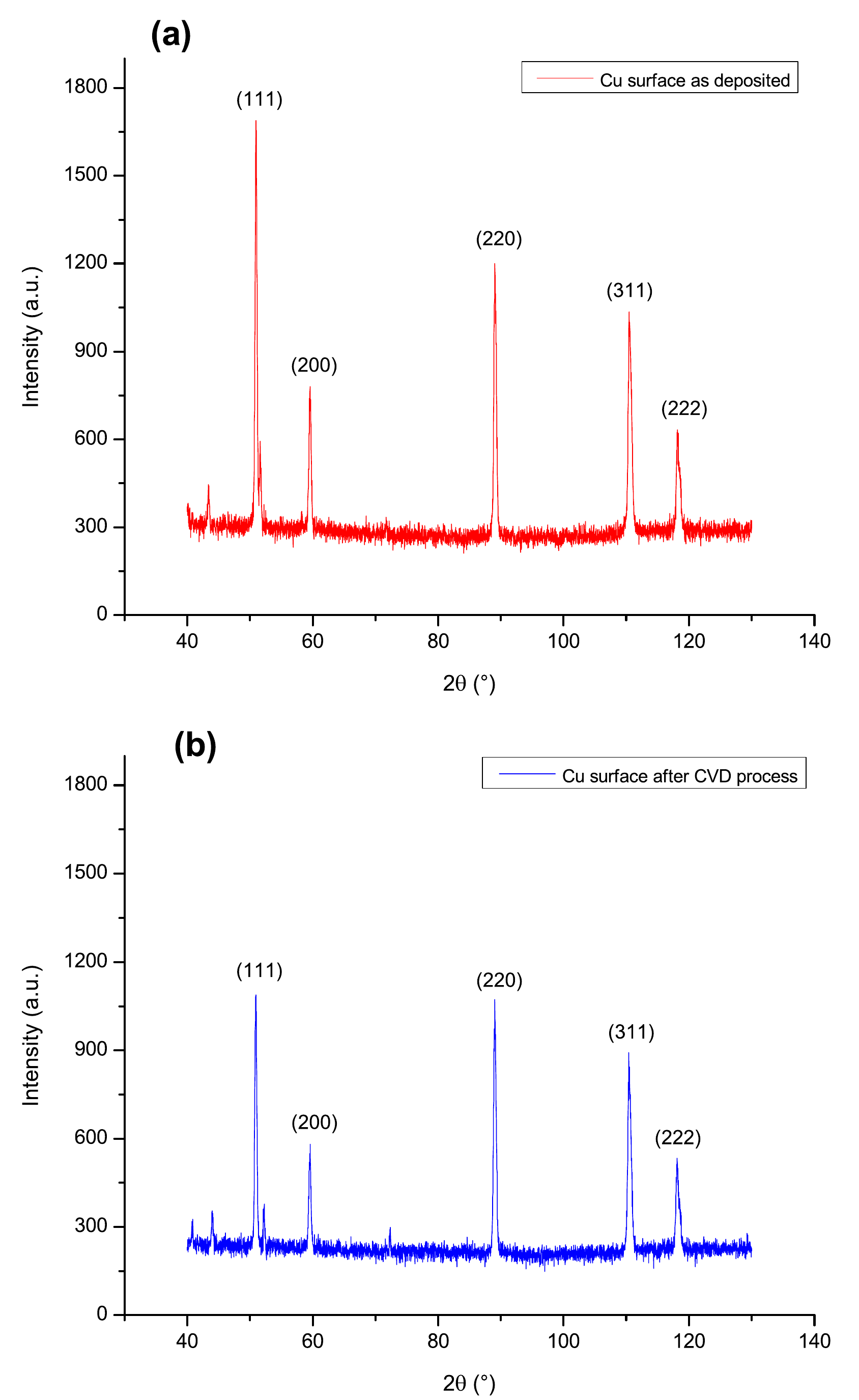}
\end{center}
\specialsubsectionfigure{\footnotesize{Figure 5 - XRD Analysis of Cu surface}}\label{XRD}
\scriptsize{XRD analysis of the Cu surface: (a) as deposited after e-beam evaporation; (b) after high temperature treatment at 725 $^\circ$C.}

\normalsize
\vskip 0.5cm
Samples have been finally characterized with Raman spectroscopy by means of a Renishaw \textit{In-Via Raman Microscope}, equipped with He-Cd blue laser at 442 nm, avoiding in this way background plasmon emission of Cu when excited e.g. by green light\cite{tao:CVD_Cufilmandfoil}. For the acquisition we have used laser power both at 1 mW and 0.1 mW, by varying consequently the exposure time of the samples to laser irradiation (in order to keep the amount of energy $E=Pt$ transferred from laser light to samples fixed).

The first spectrum (\ref{originalRaman_0.1mW800s}) has been acquired by setting the laser at low power (0.1 mW) and by exposing the sample to laser light for $t= 800$ s.

\includegraphics[keepaspectratio,scale=0.39]{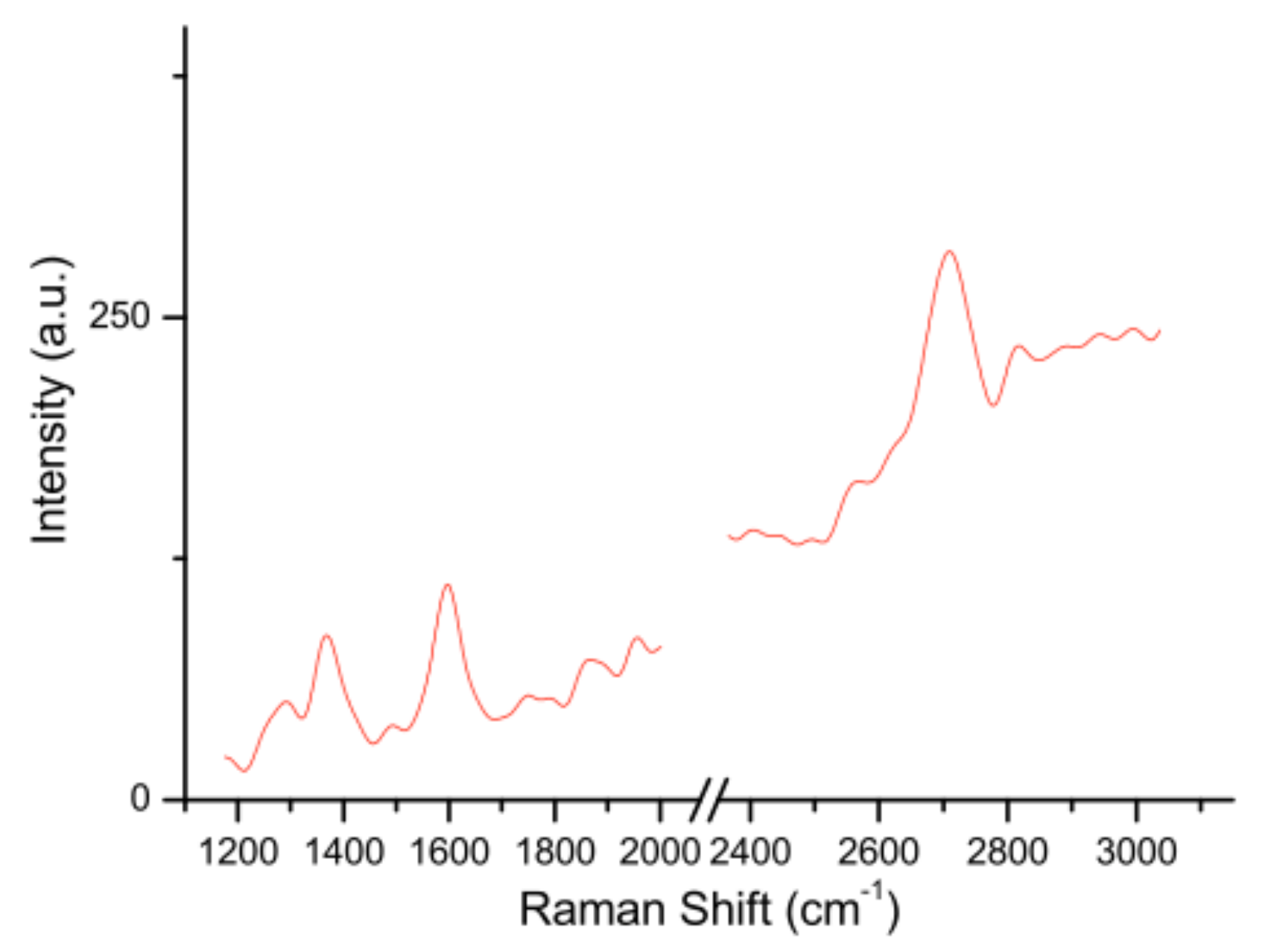}
\specialsubsectionfigure{\footnotesize{Figure 6 - Raman Spectrum of CVD graphene at low power and long acquisition time}}\label{originalRaman_0.1mW800s}
\scriptsize{First Raman spectrum of not transferred CVD graphene grown on Cu, acquired with laser power 0.1 mW for 800 s. The pronounced baseline is typical of graphene Raman spectra acquired on Cu\cite{CVDchapterbook}.}
\includegraphics[keepaspectratio,scale=0.4]{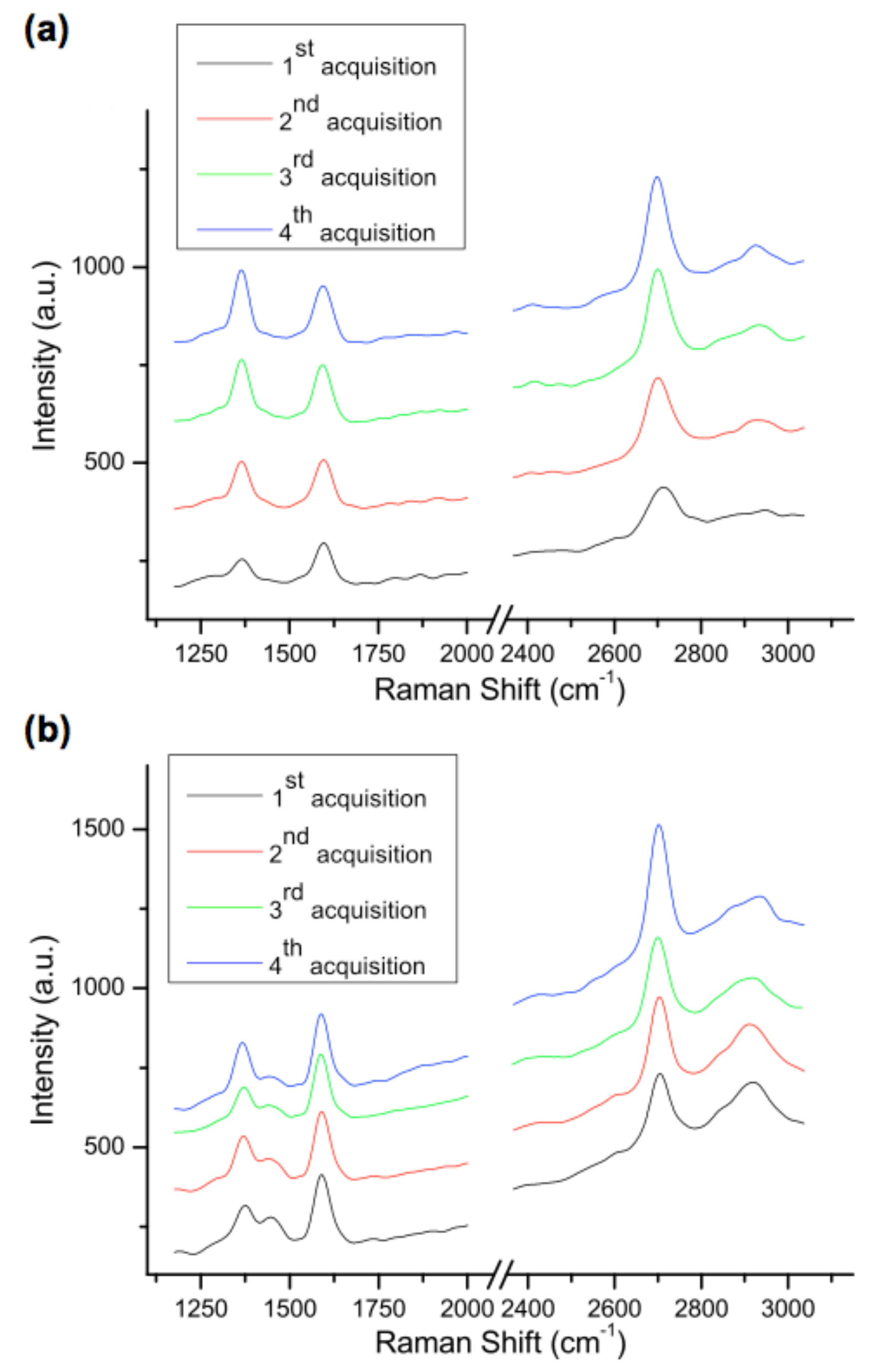}
\specialsubsectionfigure{\footnotesize{Figure 7 - Raman spectra of CVD graphene at high power and fast acquisition time}}\label{originalRaman_1mW80sec}
\scriptsize{Raman spectra of CVD graphene grown on Cu acquired on the same spot at following time intervals (all the acquisitions have been performed at 1 mW for 80 s): (a) graphene onto Cu substrate; (b) graphene transferred onto SiO$_2$ substrate.}

\normalsize
\vskip 0.5cm
By repeating four times the acquisition in the same spot (of the order of 1 $\mu$m$^2$) at higher power (1 mW) but reduced exposure time ($t=80$ s), we obtained the spectra of \ref{originalRaman_1mW80sec}(a). Three prominent peaks are visible: (a) the D peak at $\sim 1340$ cm$^{-1}$ associated to a defect induced inter-valley scattering, (b) the G peak at $\sim 1594$ cm$^{-1}$ due to in-plane optical vibrations of carbon atoms in the hexagonal crystalline structure and (c) 2D peak at a Raman shift varying between $\sim 2698$ and $\sim 2710$ cm$^{-1}$ due to a double-resonant inter-valley scattering involving two in-plane optical phonons\cite{reina:CVD,tao:CVD_Cufilmandfoil,ismach:CVD_Cu,su:CVD_Cu,malard:raman_graf}. 

It is possible to notice in particular that while the G peak shape and position are unaffected by laser exposure (only a slight increase in intensity is worthy of note), the D and 2D peaks change significantly their structure.\\ 
The careful analysis of both the position and the FWHM of the 2D peak for all spectra is reported in \ref{pos_FWHM_2dpeak_Ig/I2d_ratio} and \ref{comparison_2Dpeaks}(a).

\vskip 0.1cm
\includegraphics[keepaspectratio,scale=0.39]{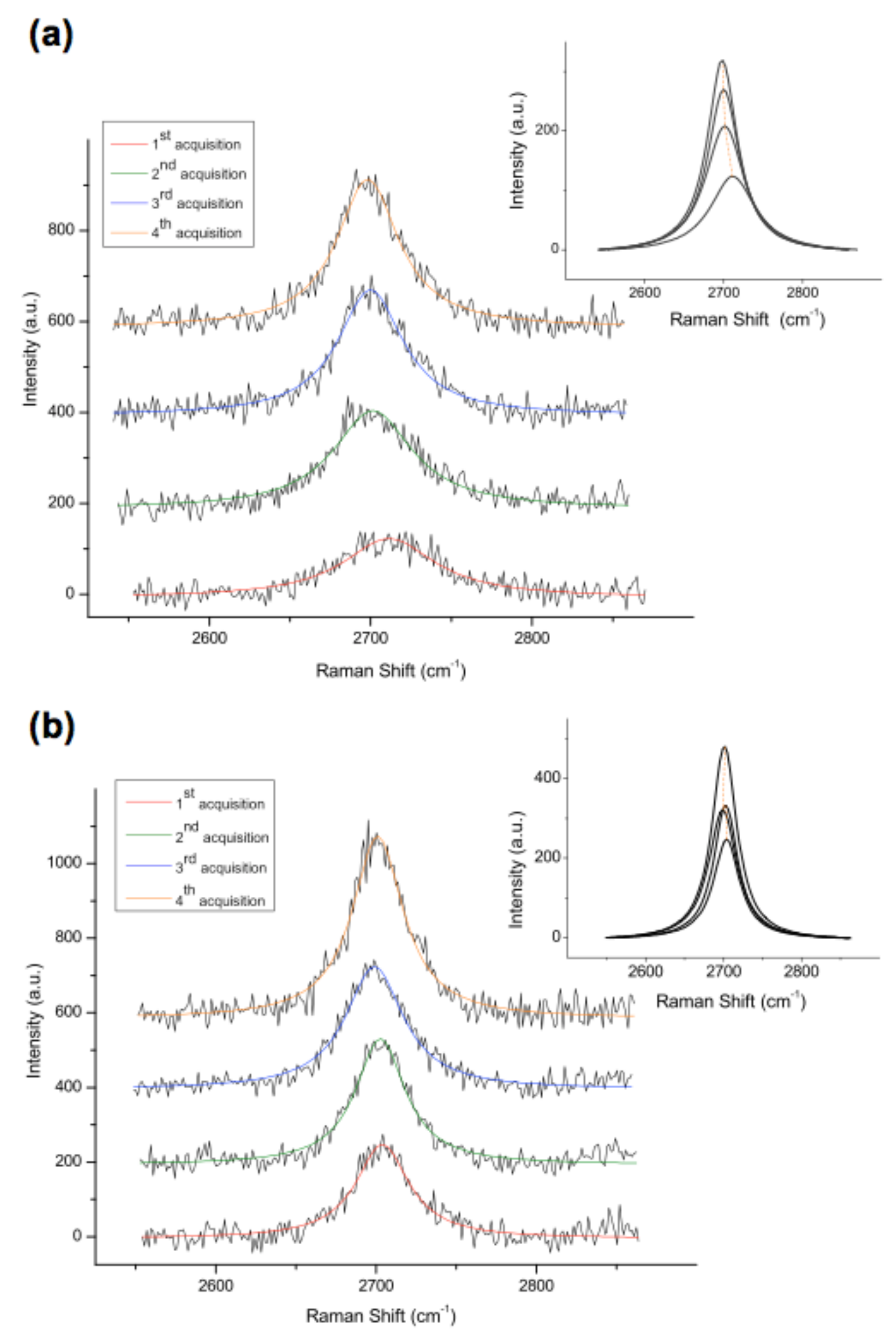}
\specialsubsectionfigure{\footnotesize{Figure 8 - 2D peak evolution with number of Raman acquisitions}}\label{comparison_2Dpeaks}
\scriptsize{Behaviour of position and shape of 2D peak of Raman spectra acquired in the same spot of the sample: (a) graphene onto Cu substrate; (b) graphene transferred onto SiO$_2$ substrate. (Insets: change in position and intensity of the single Lorentzian curves used to fit the 2D peak of the spectra).}

\normalsize
\vskip 0.5cm
The results show a significant evolution in the 2D peak shape: this fact may be attributed to some change in the graphene-like structure grown on Cu. The evolution here reported is compatible with a decrease in the number of graphene layers present on the Cu substrate, as suggested by the lowering in the Raman shift position of the peaks centre and by the sharpening of the 2D peaks observed while the exposure to laser light is increased in time. This interpretation is confirmed by the evaluation of the I$_\mathrm{G}$/I$_\mathrm{2D}$ ratio of the four spectra: the ratio is decreasing as expected for a decrease in the number of graphene layers\cite{reina:CVD,ismach:CVD_Cu,malard:raman_graf,ferrari:raman_graf}, as shown in \ref{pos_FWHM_2dpeak_Ig/I2d_ratio}. However, these results do not permit to make a clear quantitative estimation of the number of graphene layers present on the Cu substrate before and after the exposition to the laser light during Raman analysis, because they do not completely agree with the typical values reported in literature for position and FWHM of 2D peak of mono-, bi- and few-layer graphene.  

A possible phenomenological model to explain this effect, as discussed also in \cite{han:laser_etch}, is based on the observation that the laser beam at high power provides an amount of heat sufficient to locally etch away, in presence of oxygen atmosphere, outermost graphene layers grown during CVD process on top of Cu. Although it is known that suspended monolayer graphene shows a room temperature thermal conductivity of up to $\sim 5000$ W/(mK) \cite{balandin:thermal_conductivity_susp_graphene}, the effect of such a huge value can be significantly reduced in our system because of the decrease in the surface area of graphene flakes occurring while increasing number of layers (said $A^{(i)}$ the area of the surface covered by the $i$-th graphene flake and $A^{(\mathrm{Cu})}$ the area of the underlying Cu grain, $A^{(n)}>A^{(n-1)}>\ldots>A^{(2)}>A^{(1)}>A^{(\mathrm{Cu})}$). Moreover, Raman mapping shows different regions with different number of layers on the same sample, corroborating our hypothesis of a terraces-like structure characterizing all the substrate. This fact implies that in-plane heat dissipation through outer graphene layers is largely suppressed by finite-size effects and out-of-plane heat transfer (much lower than the previous one) becomes the dominant heat transfer channel, resulting in a local overheating and subsequent etching of the layers (the model is schematically represented in \ref{phenomenological_model_heating}). On the contrary, the innermost graphene layer directly bounded to the Cu surface is protected from this effect by the presence of the substrate acting as heat sink. Indeed, it has been reported\cite{ruoff:thermal_conductivity_on_Cu} that Cu lowers the thermal conductivity of CVD monolayer graphene grown on it: a significant change in the enhancement of the temperature and in the G peak shifting as a function of the absorbed laser power by using Cu instead of SiO$_2$ as a substrate has been reported in this work. As a consequence, graphene-on-Cu and Cu systems have comparable thermal conductivities and the effect of the substrate is not anymore negligible: being Cu a good heat conductor, it provides in turn the dominant contribution to heat dissipation and allows for a more efficient heat exchange through the substrate with respect to the case of transferred graphene onto SiO$_2$. 

At such photon energies other mechanisms, like molecular desorption of chemical species, cannot be \textit{a priori} excluded, but they can hardly affect the Raman signature and they can be detected through electrical measurements\cite{sun:photoinduced_desorption}.

\begin{center}
\includegraphics[keepaspectratio,scale=0.4]{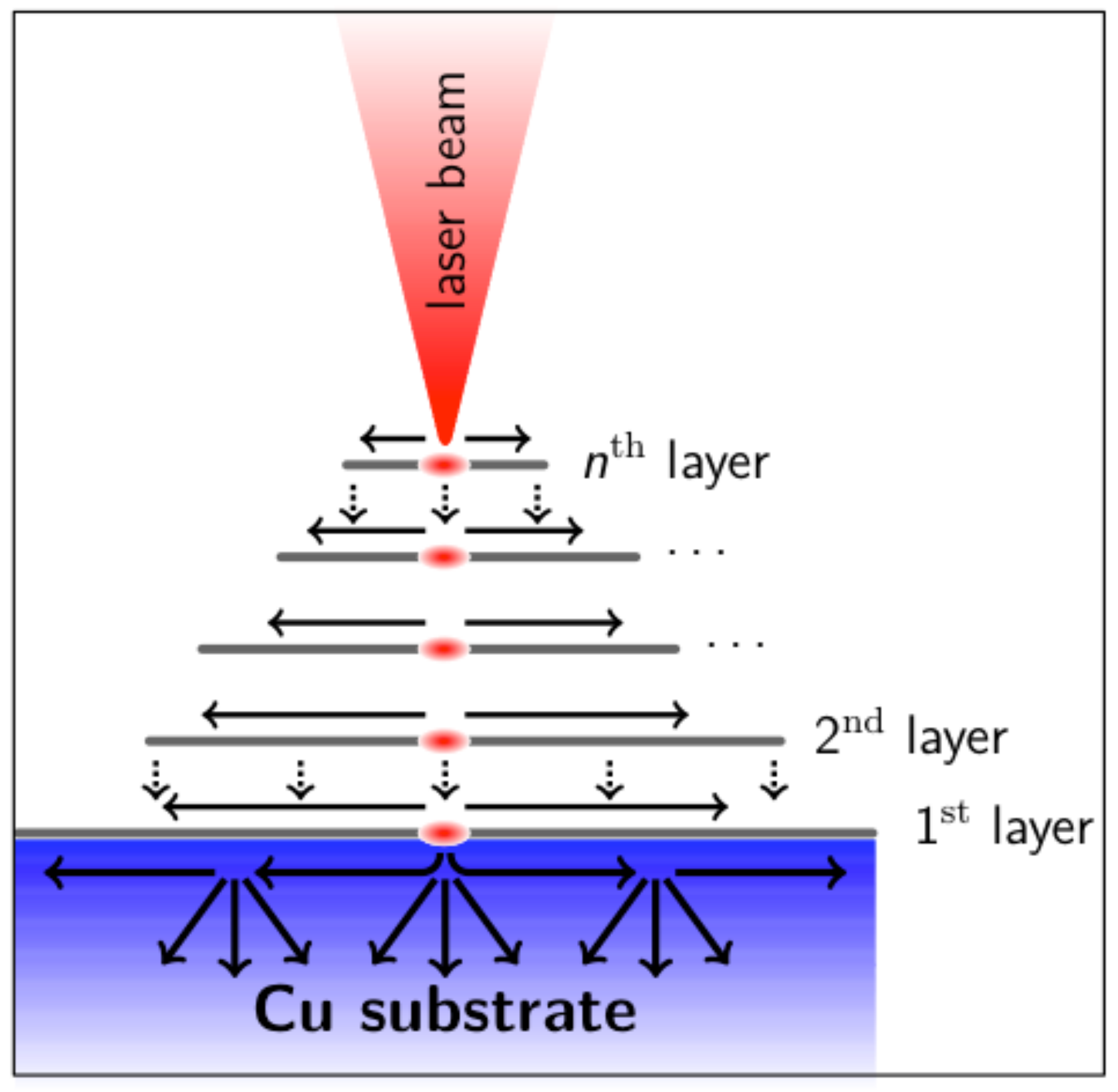}
\end{center}
\specialsubsectionfigure{\footnotesize{Figure 9 - Phenomenological model for local overheating and etching of multilayer graphene}}\label{phenomenological_model_heating}  
\scriptsize{Schematic model representing the possible origin of local overheating and etching of outermost graphene layers grown on Cu films. The heat provided by the focused laser beam is dissipated through in-plane (horizontal arrows) and out-of-plane (vertical arrows) channels. Though the in-plane channel is dominant in graphene, the finite size of outer layers makes out-of-plane dissipation significant and reduces the in-plane contribution, thus causing local overheating and subsequent etching of the layers. The innermost layer is instead prevented from etching by the presence of the Cu substrate acting as heat sink.}

\normalsize
\vskip 0.5cm
The fact that the G peak position is almost unaffected by the laser irradiation can be explained as a result of two competing effects: the local enhancement of temperature (due to overheating), bringing to a redshift of the G peak\cite{ruoff:thermal_conductivity_on_Cu,calizo:Gpeak_temperature} and the decrease in the number of graphene layers (due to etching), resulting instead in a blueshift of the G peak\cite{wang:Gpeak_numberlayers}.\\ 
The unexpected unaltered intensity of the G peak (that should decrease as the number of graphene layers decreases) can be explained in two ways. A first effect, applying to Raman measurements performed both on transferred and not transferred graphene samples, relies on the increase in temperature of graphene layers upon laser irradiation, likely resulting in an effect similar to what observed in experiments concerning the evolution of graphene Raman signature upon controlled annealing at high temperatures\cite{ni:annealing}. As pointed out in this work, the G peak intensity is not changing between few- and monolayer graphene sheets after the annealing process, meaning that in these experimental conditions G peak intensity cannot be regarded as a fingerprint to distinguish number of graphene layers. A second reasoning, applying only on graphene samples over Cu, relies on the roughness of the Cu surface, determining a light trapping effect close to the substrate's surface that results in an enhanced number of multiple reflections of the laser light between Cu and graphene layers: as a consequence, the number of C atoms detected by the unfocused beam is always comparable to the number of C atoms present in multilayer graphene, although the number of graphene layers is decreasing. As reported in\cite{ni:imaging_graphene} the G band intensity for a number of graphene layers exceeding $\sim 15$ is in this case decreasing by increasing number of layers and then constant, as observed in our spectra.

The origin of the prominent D peak (increasing in intensity as the number of acquisitions increases) is not completely clear yet: it can be attributed to the acquisition of the spectra on a point of the sample lying on a grain boundary of the Cu substrate (resulting in a change of the crystallographic orientation of graphene flake through it) or to defects (terrace boundaries) produced in graphene crystal structure by the laser etching.

Raman analysis performed on transferred graphene samples using the same experimental setup (laser at 442 nm, power $P=1$ mW and exposure time $t=80$ s) confirms the behaviour observed in the spectra acquired on graphene/Cu substrates. A similar evolution in shape and positions of the peaks is obtained for subsequent spectrum acquisitions, as shown in \ref{originalRaman_1mW80sec}(b) and \ref{comparison_2Dpeaks}(b).

\section*{Conclusions}
In summary, a possible way to locally etch graphene layers based on laser heating released during Raman analysis, has been presented. Graphene structure (crystallization degree and number of layers) evolution can be monitored and inferred by looking at the Raman spectra acquired on the sample. The technique is suitable in particular for etching layers of graphene grown by CVD on a metal catalyst. In our case we have reported results obtained with Cu: by lightening the sample with incident laser light at quite high power ($\gtrsim 1$ mW) for short time periods ($\sim 80$ s) a clear  sharpening and lowering of the 2D peak position is observed in Raman spectra, together with a decrease in the I$_\mathrm{G}$/I$_\mathrm{2D}$ ratio. These results are compatible with a decrease in the number of graphene layers grown on the metallic substrate. D peak increases in intensity as function of the laser exposure, meaning an increasing of defects in the graphene structure.\\ We believe that the method can be easily applied to other metallic substrates: since the technique deeply relies on the dispersion of the heat provided by laser irradiation through the substrate, the most important feature required for the substrate is its high thermal conductivity.   

Although some catalysts, like Cu, are very promising for CVD synthesis of graphene because of the expected self-limited mono layered growth of this material on them, a few-layered structure is often found in experiments. Laser etching here reported can therefore provide an \textit{in-situ} technique to get rid of this problem. However, laser can also have an active r\^{o}le in inducing unwanted defects, such as vacancies in pristine graphene films: for this reason the proposed method shall be further developed. We finally envisage the application of the photo-etching process here reported to large areas if efficient and uniform illumination conditions, as those used in our RTA system, are employed.

\bigskip

\section*{List of abbreviations}
   Abbreviations used in the text:
    \begin{itemize}
     \item CVD: Chemical Vapour Deposition;
     \item SEM: Scanning Electron Microscopy;
     \item STM: Scanning Tunneling Microscopy;
     \item RMS: root mean square;
     \item AFM: Atomic Force Microscopy.
     \item RTA: Rapid Thermal Annealing;
     \item PMMA: Poly(methyl-methacrylate);
     \item XRD: X-Ray Diffraction.
    \end{itemize}

\subsection*{Author's contributions}
     MP has been responsible for data analysis and modeling and for Cu e-beam evaporation, helping also in the theoretical interpretation of the results. LC coordinated all the experimental work, in particular the Cu e-beam evaporation and the CVD process, participating also in Raman analysis. EV and GA coordinated all the work and have been responsible for the theoretical interpretation of the results.

\subsection*{Competing interests}
  \ifthenelse{\boolean{publ}}{\small}{}
  The authors declare that they have no competing interests.
  
\section*{Acknowledgements}
  \ifthenelse{\boolean{publ}}{\small}{}
  The authors acknowledge the Nanofacility Piemonte and the Compagnia di San Paolo for financial supports. Thin films evaporation, graphene deposition and SEM analysis on the samples have been performed at Quantum Laboratories - Nanofacility Piemonte present at I.N.Ri.M. - Turin. Raman characterization has been performed with the help of A. Damin at the NIS Centre of Excellence - University of Turin. The help of A. Battiato and E. Olivetti with XPS and XRD analyses is gratefully acknowledge. 
  

\newpage
{\ifthenelse{\boolean{publ}}{\footnotesize}{\small}
 \bibliographystyle{bmc_article}  
  \bibliography{article_laser_ablation_REVISED} }     


\ifthenelse{\boolean{publ}}{\end{multicols}}{}

\section*{Tables}
  \specialsubsectiontable{Table 1 - 2D peak position and sharpness, I$_{\boldsymbol{\mathrm{G}}}$/I$_{\boldsymbol{\mathrm{2D}}}$ ratio}\label{pos_FWHM_2dpeak_Ig/I2d_ratio}
    Comparison of the position and FWHM of 2D peaks of subsequent Raman acquisitions as extracted by a single Lorentzian fit and evolution of the I$_\mathrm{G}$/I$_\mathrm{2D}$ ratio (spectra from \ref{originalRaman_1mW80sec}, \ref{comparison_2Dpeaks}). Uncertainties of the order of $0.01 - 0.04\%$ for the 2D peak Raman shifts and of $3 - 6\%$ for 2D peak FHWMs have been estimated. Note: results concerning transferred graphene are affected by a higher uncertainty because of pronounced secondary peaks (possibly due to residual PMMA used for graphene transfer) present in the Raman spectra (see \ref{originalRaman_1mW80sec}(b)), making the fit procedure less accurate. \par \mbox{}
    \par
    \mbox{
      \begin{tabular}{*{7}{c}}
\toprule
\bfseries{\multirow{3}*{Acquisition}} & \multicolumn{3}{c}{\bfseries{\emph{Graphene on Cu}}} & \multicolumn{3}{c}{\bfseries{\emph{Graphene on SiO$_2$}}} \\
 & \bfseries{Raman Shift} & \bfseries{FWHM} & \bfseries{\multirow{2}*{I$_{\boldsymbol{\mathrm{G}}}$/I$_{\boldsymbol{\mathrm{2D}}}$}} & \bfseries{Raman Shift} & \bfseries{FWHM} & \bfseries{\multirow{2}*{I$_{\boldsymbol{\mathrm{G}}}$/I$_{\boldsymbol{\mathrm{2D}}}$}}\\
& \bfseries{(cm$^{\boldsymbol{-1}}$)} & \bfseries{(cm$^{\boldsymbol{-1}}$)} & & \bfseries{(cm$^{\boldsymbol{-1}}$)} & \bfseries{(cm$^{\boldsymbol{-1}}$)} & \\
\midrule
1 & 2711.4 & 70.0 & 0.86 & 2704.0 & 40.3 & 1.01\\
\midrule
2 & 2701.3 & 60.0 & 0.60 & 2702.7 & 37.7 & 0.79\\
\midrule
3 & 2699.8 & 49.2 & 0.56 & 2699.0 & 43.8 & 0.80\\
\midrule
4 & 2698.2 & 46.6 & 0.46 & 2701.1 & 38.7 & 0.61\\
\bottomrule
\end{tabular}
}

\end{bmcformat}
\end{document}